\documentclass[twocolumn, twocolappendix]{aastex631}

\newcommand{\orcid}[1]{\href{https://orcid.org/#1}
{\includegraphics[width=8pt]{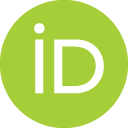}}}

\usepackage{amsmath}

\begin{document}

\title{Investigating potential causes for high inferred intrinsic temperatures of warm transiting Neptune and Sub-Neptune planets}

\correspondingauthor{Saugata Barat}
\email{saugatabarat500@gmail.com}

\author[0009-0000-6113-0157]{Saugata Barat}
\affiliation{Department of Astronomy, University of Texas, Austin, TX 78712, USA}

\affiliation{Kavli Institute for Astrophysics and Space Research, Massachusetts Institute of Technology, Cambridge, MA 02139, USA}

\author[0000-0002-0875-8401]{Jean-Michel D\'esert}

\affiliation{Leibniz Institute for Astrophysics Potsdam, An der Sternwarte 16, 14482 Potsdam, Germany}

\affiliation{Anton Pannekoek Institute for Astronomy, University of Amsterdam,
Science Park 904, 1098 XH,
Amsterdam, the Netherlands}

\author[0000-0001-9504-3174]{Allona Vazan}
\affiliation{Astrophysics Research Center (ARCO), Department of Natural Sciences, 
The Open University of Israel, 
Ra’anana, 43107, Israel}

\author[0000-0002-6576-3346]{Ritika Sethi}
\affiliation{Department of Physics, Massachusetts Institute of Technology, Cambridge, MA 02139, USA}
\affiliation{Kavli Institute for Astrophysics and Space Research, Massachusetts Institute of Technology, Cambridge, MA 02139, USA}

\author[0000-0003-1622-1302]{Sagnick Mukherjee}
\affiliation{School of Earth and Space Exploration, Arizona State University, Tempe, AZ, USA}

\author[0000-0003-3130-2282]{Sarah Millholland}
\affiliation{Department of Physics, Massachusetts Institute of Technology, Cambridge, MA 02139, USA}
\affiliation{Kavli Institute for Astrophysics and Space Research, Massachusetts Institute of Technology, Cambridge, MA 02139, USA}

\author[0000-0000-0000-0000]{Jonathan J. Fortney}
\affiliation{Department of Astronomy and Astrophysics, 
University of California, 
Santa Cruz, 95064, CA, USA}

\author[0000-0001-6315-7118]{William Misener}
\affiliation{Earth and Planets Laboratory, Carnegie Institution for Science, 5241 Broad Branch Road NW, Washington, DC 20015, USA}

\author[0000-0002-0298-8089]{Hilke E. Schlichting}
\affiliation{Department of Earth, Planetary, and Space Sciences, The University of California, Los Angeles, 595 Charles E. Young Drive East, Los Angeles, CA 90095, USA}

\author[0000-0002-0298-8089]{Sara Seager}
\affiliation{Department of Earth, Atmospheric and Planetary Sciences, Massachusetts Institute of
Technology, 77 Massachusetts Avenue, Cambridge, MA 02139, United States}



\begin{abstract}

Atmospheric characterization of a warm 10-30~Myr old sub-Neptune progenitor (V1298 Tau b) and a $\sim$3.4~Gyr old warm super-Neptune (WASP-107b) using JWST have revealed extremely low methane abundance. Atmospheric forward models find higher than expected inferred intrinsic temperatures (450~K and 500~K) --- inconsistent with theoretical planet formation and core-envelope evolution predictions (70~K and 150~K respectively) assuming a convective interior. We explore three hypotheses to reconcile the high intrinsic temperatures with evolutionary models --- tidal heating, silicate rainout and heat trapping due to deep cloud condensation. Tidal heating requires high planetary obliquity ($\gtrsim70^{\circ}$) and low reduced tidal quality factor ($\sim$100) for V1298 Tau b, implying an extremely short obliquity damping timescale ($<$10$^{4}$ years). Although high obliquity may be maintained by spin--orbit resonant locking, it requires fine-tuning and is therefore unlikely. Silicate rainout can increase the temperature in the deep atmosphere of the young V1298 Tau b (0.1-1~bar) by $\sim$700~K, without requiring extremely high intrinsic temperature, but is unlikely to explain WASP-107b, as silicate rainout timescales are expected to be much shorter than its age. Including cloud condensation in self-consistent atmospheric models reduces the inferred intrinsic temperature for V1298 Tau b from 500~K to 300~K, still exceeding the core-envelope prediction (150~K). Future measurements of thermal emission from young transiting planets such as V1298 Tau b, combined with methane abundances as a function of age, will be needed to distinguish between these scenarios.

\end{abstract}



\section{Introduction}


Understanding the formation and evolution of these planets requires us to understand their interior structure \citep{rogers2010,thorngren2019,lichtenberg2025} and entropy \citep{fortney2010,marley2007}. Transmission spectroscopy typically probes high up in the atmosphere ($\sim$1~mbar) and is not directly sensitive to the interior structure and temperature. However, it has been proposed that methane (CH$_4$) abundance in the upper atmosphere could be used as a probe of thermal conditions in the deep atmosphere ($>$0.1-1 bar) \citep{fortney_2020} --- the so-called `methane thermometer'. For warm planets($T_{eq} <$800~K) the dominant C-bearing molecule is CH$_4$, while for hotter planets CO/CO$_2$ dominates the carbon budget \citep{lodders2002,MosesEtal2011apjDisequilibrium}. However, it has been shown that vertical mixing can remove methane from the observable atmosphere of warm planets by quenching it at high pressures \citep{zahnle2014,Baxter2021,mukherjee2024}. Therefore, the observable methane abundance is a tracer for thermal conditions in the deep atmosphere.


Using the `methane thermometer' concept, the temperature of the deep atmosphere ($\sim$ 0.1-1 bar) was inferred for two warm transiting planets --- the super-Neptune WASP-107b (10.5~R$_{\oplus}$, 30.5~M$_{\oplus}$, 3.4 Gyr old, \citealt{anderson2017,piaulet2021}) and the young sub-Neptune progenitor V1298 Tau b (9.4~R$_{\oplus}$, 12~M$_{\oplus}$, 23~Myr old, \citealt{david2019,livingston2026,barat2025}). In Table \ref{tab:table1} we compare the properties of both these planets.

\begin{figure*}
    \centering
    \includegraphics[width=1\textwidth]{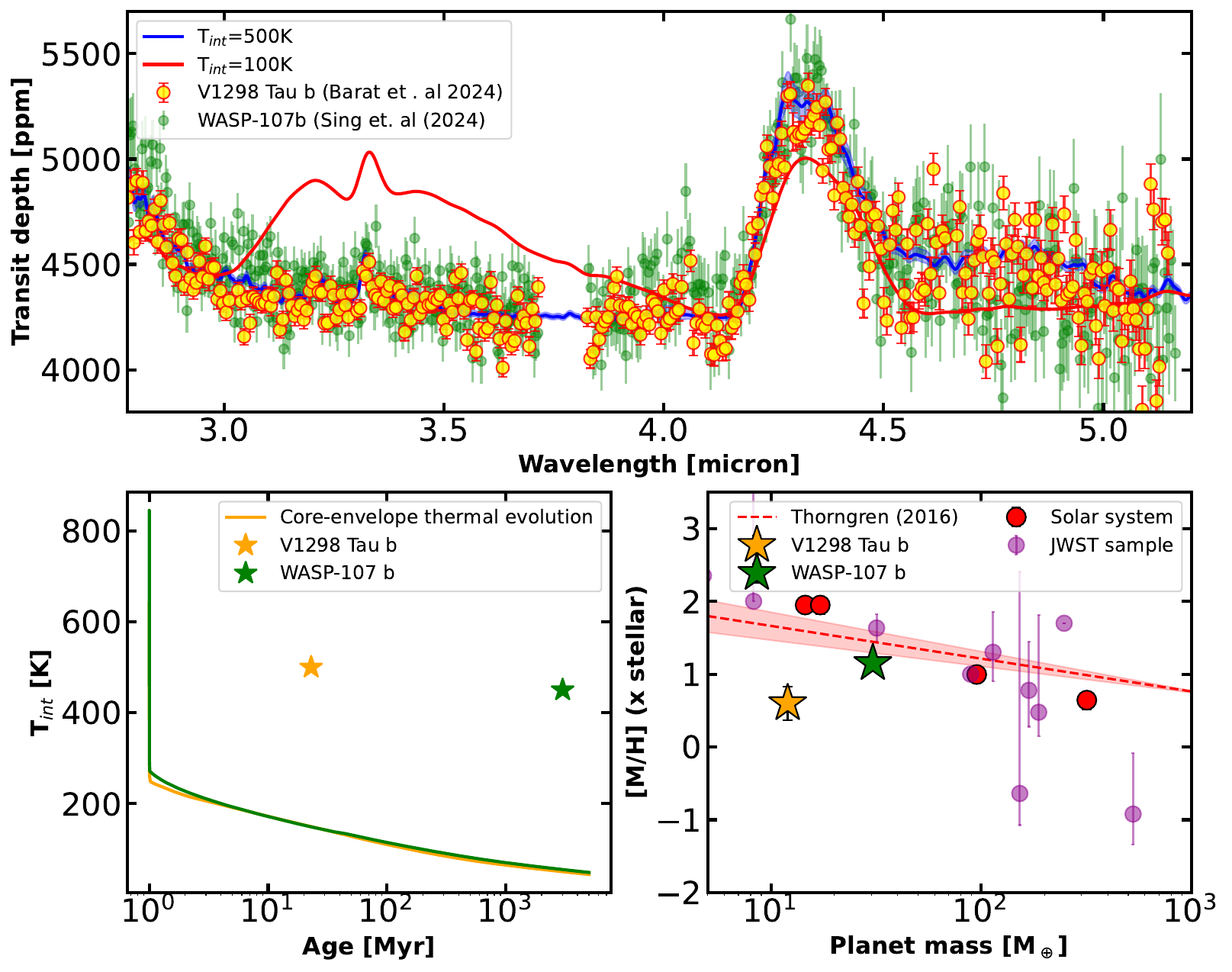}
    \caption{Upper panel: Comparing the JWST NIRSpec G395H transmission spectra of V1298 Tau b (yellow) and WASP-107b (green) taken from \citet{barat2025} and  \citet{sing2024} respectively. WASP-107b spectra have offset by 10000 ppm for the ease of visual comparison. Blue solid line shows a PICASO self-consistent forward model for V1298 Tau b for T$_{int}$ of 500~K and K$_{zz}$=10$^7$. Red solid line shows a model for V1298 Tau b with similar parameters (mass, isothermal temperature, metallicity, C/O ratio and K$_{zz}$), but a T$_{int}$ of 100~K. These models assume a convective interior. Low intrinsic temperature model produces a large methane feature around 3.3~$\mu$m and can be ruled out from the observed spectra of both planets. Lower left: Thermal evolution models for V1298 Tau b and WASP-107b simulated using core-envelope interior structure for V1298 Tau b and WASP-107b following formalism outlined in \citet{vazan2018}. The parameters at the current age used to benchmark our evolution models are provided in Table \ref{tab:table1}. The expected intrinsic temperature at the current age is$\sim$150~K and 100~K for V1298 Tau b and WASP-107b respectively. Lower right: Mass metallicity diagram showing published atmospheric metallicities from JWST observations taken from \citet{barat2024b}. We overplot V1298 Tau b \citep{barat2025} and WASP-107b \citep{sing2024,welbanks2024}. We also show the theoretical mass-metallicity trend derived for exoplanets from \citet{thorngren2019} assuming a core accretion formation model.}
    \label{fig:figure1}
\end{figure*}

JWST NIRSpec and NIRCam observations of WASP-107b detected methane, but at a much lower abundance compared to equilibrium chemistry prediction \citep{sing2024,welbanks2024}. Similarly for V1298 Tau b, a combination of HST/WFC3 and JWST NIRSpec G395H observations detected CH$_4$ but at a relatively low abundance: two orders of magnitude lower than equilibrium chemistry prediction \citep{barat2024,barat2025}. In Figure \ref{fig:figure1} we compare the NIRSpec/G395H transmission spectra of both these planets, which appear remarkably similar. The lack of methane has been reported in other planets as well, including GJ3470b \citep{beaty2024} and V1298 Tau e \citep{dai2026}. This mystery of the `missing methane problem' is emerging as a problem for different planetary systems across mass, age and temperature \citep{yu2026}.

If we assume that the deep atmosphere is adiabatic (i.e efficient convection in the interior), to explain the high temperature required to reproduce the observed methane abundance in the deep atmosphere, we need very high intrinsic luminosity (T$_{int}$) --- 450~K for WASP-107b \citep{sing2024,welbanks2024} and 500~K for V1298 Tau b \citep{barat2025}. The intrinsic temperature (T$_{int}$) is a blackbody temperature assumed to parameterize the intrinsic luminosity of the planet. In Figure \ref{fig:figure1} we compare the inferred intrinsic temperature for both planets with predictions from thermal evolution models. The inferred intrinsic luminosity is $\sim$150 and 400$\times$ higher than the prediction from thermal evolution models for V1298 Tau b and WASP-107b respectively. Therefore, additional sources of energy are necessary to explain these observations.

For WASP-107b tidal heating has been proposed \citep{millholand_2020, welbanks2024, 2025ApJ...988..247S}. However, follow-up radial velocity measurements have challenged this hypothesis \citep{yee2026}. For V1298 Tau b, the presence of frequent flaring has been proposed to remove methane due to photolysis \citep{adams2026}, although it depends sensitively on the assumed energy distribution of the injected flares in their models.


Therefore, there is a growing need for an explanation of the `missing methane/hot interior' problem for exoplanets. In this paper we explore and evaluate three potential explanations --- i. enhanced tidal heating due to obliquity tides \citep{millholand_2020} ii. non-adiabatic interiors due to silicate rainout in the envelope leading to slower cooling \citep{vazan18b,vazan2024} and iii. deep cloud condensation resulting in heat trapping \citep{mukherjee2026}. We base our calculations on V1298 Tau b. We briefly describe the V1298 Tau system properties in Section \ref{v1298tau}. In Section \ref{results} we outline the three models and discuss the key findings. In Section \ref{discussion} we discuss and interpret the results and future directions and summarize our findings in Section \ref{conclusion}.

\begin{table*}
    \centering
    \caption{Comparing the physical properties of V1298 Tau b and WASP-107b.}
   
    \begin{tabular}{c|c|c}
    \hline\hline
   
      Planet   & V1298 Tau b \footnote{\citet{david2019b}} & WASP 107b \footnote{\citet{anderson2017}} \\
      Radius [R$_{\oplus}$]   & 9.4$\pm$0.6 & 10.4$\pm$0.1 \\
      Orbital distance [AU]   &  0.16 &  0.05\\
      Equilibrium Temperature [K]   & 670$\pm$20 & 770$\pm$60\\
      Mass [M$_{\oplus}$]   &  12$\pm$1 \footnote{\citet{barat2025}} & 30$\pm$1.5 \footnote{\citet{piaulet2021}}\\
      Density (g/cm$^3$ & 0.06$\pm$0.01 & 0.13$\pm$0.01 \\
      logZ/Z$_{\odot}$ &  0.6$^{+0.2}_{-0.3}$ \footnote{\citet{barat2025}} & 1.2$^{+0.1}_{-0.2}$ \footnote{\citet{welbanks2024}} \\
      C/O ratio &  0.22$\pm$0.02  & 0.33$\pm$0.06 \\
      Methane detection [$\sigma$] & 6 & 5 \\
      T$_{int}$ [K] & 500 & 450 \\
      K$_{zz}$ & $<$10$^{7}$ & 10$^{8.4}$-10$^{9}$ \\
      \hline \hline
      Star & V1298 Tau & WASP-107 \\

      Age [Gyr]  & 10-30~Myr & 3.4$\pm$0.4\\
      Spectral Type & K1 &   K6\\
      J mag & 8.7 & 9.4\\
       

    \hline
    \end{tabular}

    \label{tab:table1}
\end{table*}

\section{V1298 Tau system} \label{v1298tau}

V1298 Tau is a young multi-planet system (10-30~Myr) with four transiting planets (c,d,b and e) with orbital periods of 8.24, 12.40, 24.14 and 48.67 days \citep{david2019b,livingston2026}. Different studies have reported slightly different ages --- 12$\pm$2~Myr \citep{finoceti2023}, 20$\pm$1~Myr \citep{mascareno_2021}, 23$\pm$4~Myr \citep{david2019} and 28$\pm$4~Myr \citep{johnson2022}. However, for this paper we have adopted the 23~Myr age estimate which does bear significant effect on the overall conclusion. The planets c,d,b and e have 5.08$\pm$0.37, 6.53$\pm$0.42, 9.41$\pm$0.57 and 10.17$\pm$0.75~R$_{\oplus}$ respectively. The four planets are found to exhibit strong TTVs, yielding masses of 4.7$\pm$0.6, 6$\pm$0.7, 13.1$\pm$5.3 and 15.3$\pm$4.2~M$_{\oplus}$ and equilibrium temperatures of 950, 850, 670 and 520~K for planets respectively \citep{livingston2026}. JWST observations of three planets (b, c and e) in the system reveal mass constraints consistent within 2$\sigma$ of the TTV estimates \citep{barat2025,dai2026,murphy2026}, and atmospheric metallicities between 5-15$\times$solar. For V1298 Tau b, which we use for this paper has a gas-to-core ratio of 0.2-8\% depending on the assumption of its intrinsic luminosity \citep{barat2025}. The TTV models suggest that all four planets have relatively low eccentricity ($<$0.0094, $<$0.0087, 0.0079$\pm$0.0041 and $<$0.0124 \citealt{livingston2026}).

\section{Possible Scenarios Leading to High Intrinsic Temperature} \label{results}

\subsection{Tidal heating in V1298 Tau b} \label{tidal heating}
Tidal forces drive systems toward orbital circularization \citep{1980A&A....92..167H, 2008ApJ...678.1396J, 2023MNRAS.525..876M}, spin synchronization \citep{2007ApJ...661.1180O, 2008Icar..193..641L, 2024MNRAS.529.4442S, 2025ApJ...990..124H, 2026MNRAS.tmp..421S}, and spin–orbit alignment \citep{2009MNRAS.395.2268B, 2026MNRAS.545f2236S}. When these equilibrium states are not satisfied, the star induces a time-varying tidal deformation on the planet, dissipating orbital and/or spin energy as heat \citep{2008ApJ...678.1396J,delisle2014}. This process is known as tidal heating and it can enhance the intrinsic temperature of the planet. In particular, for a young planet like V1298 Tau b, tidal heating may possibly be ongoing with circularization timescales $\gtrsim$100~Myr \citep{Seager2002} and could provide an additional external source of energy.

\subsubsection{Required tidal luminosity and obliquity damping}
We quantify this heating using the tidal luminosity, $L_{\rm tide}$, defined as the rate at which the orbital energy is dissipated as heat within the planet \citep{1980A&A....92..167H}, using V1298 Tau b as a test case. We compute $L_{\rm tide}$ using Eq.~13 of \citet{2010A&A...516A..64L}, which uses the equilibrium tide model with constant time lag. Details of our calculation are given in Appendix~\ref{app:tides}.


\begin{figure} 
    \centering
    \includegraphics[width=\linewidth]{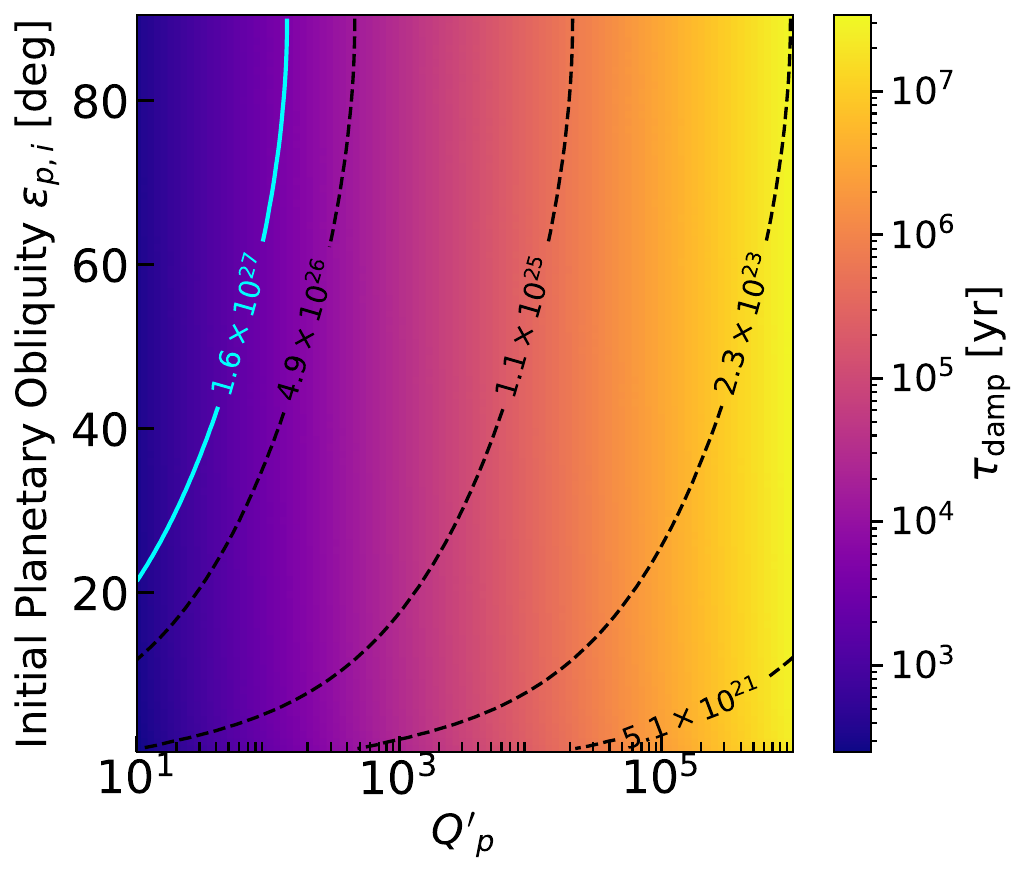}
    \caption{The obliquity damping timescale, $\tau_{\rm damp}$, as a function of $Q'_p$ and the initial obliquity $\epsilon_i$. Overplotted contours show the tidal luminosity, $L_{\rm tide}$, computed at a fixed eccentricity $e=0.01$ and evaluated across the full range of initial obliquities shown. The cyan contour corresponds to $L_{\rm tide} = L_{\rm tide, req}$.}
    \label{fig:Qp_vsep_meshplot}
\end{figure}

To investigate whether tidal heating can explain the high inferred $T_{\rm int}$ (assuming an convective interior) of V1298 Tau b, we assume the extreme case for which all excess luminosity (not accounted for by thermal evolution models) is supplied by tides. Treating intrinsic emission as blackbody radiation, we estimate the required tidal luminosity as $L_{\rm tide, req} = 4\pi R_p^2\sigma_{\rm SB}(T_{\rm int,inferred}^4-T_{\rm int,model}^4)$, where $R_p$ is the planet's radius, $\sigma_{\rm SB}$ is the Stefan-Boltzmann constant, $T_{\rm int,inferred}$ is the inferred intrinsic temperature and $T_{\rm int,model}$ is the value predicted by thermal evolution models (Figure \ref{fig:figure1}). We assume  T$_{\rm int,model}^4)$=150~K (Figure \ref{fig:figure1}) and T$_{\rm int,inferred}^4$=500~K. This gives $L_{\rm tide,req}\approx 1.6\times10^{27}$ erg s$^{-1}$. 


We first evaluate whether this luminosity can be supplied by eccentricity tides alone. V1298 Tau b has a low eccentricity, $e = 0.0079 \pm 0.0041$ \citep{livingston2026}. The reduced tidal quality factor $Q'_p$ is highly uncertain; if it is similar to Uranus and Neptune in the Solar System, then it would be $\sim10^4$--$10^5$ \citep{1999ssd..book.....M, 1990Icar...85..394T, 2008Icar..193..267Z}. To obtain a conservative upper limit on $L_{\rm tide}$ and to allow for the possibility that young planets may have lower $Q'_p$, we adopt an extremely low value, $Q'_p\approx10^2$. Even in this optimistic case, eccentricity tides alone, evaluated at planetary obliquity ($\epsilon_p$)=0$^\circ$, produce only $L_{\rm tide} \approx 7.8\times10^{16}$ erg s$^{-1}$, nearly ten orders of magnitude below $L_{\rm tide,req}$. The luminosity is calculated using the formulation outlined in \citet{millholland2019}.

While the planetary obliquity $\epsilon_p$ remains unconstrained, young planets may possess large axial tilts. Therefore, including the contribution from obliquity tides can, in principle, yield higher $L_{\rm tide}$ values. In Figure~\ref{fig:Qp_vsep_meshplot} shows the contours of $L_{\rm tide}$ in the $Q'_p$--$\epsilon_p$ plane. Only a small region in parameter space, enclosed within the cyan contour, satisfies $L_{\rm tide}\gtrsim1.6\times10^{27}$ erg s$^{-1}$, corresponding to $Q'_p\lesssim10^2$ and $\epsilon_p\gtrsim70^\circ$. Even if such low $Q'_p$ values are achievable for young planets, the associated strong tidal dissipation could rapidly damp $\epsilon_p$. 


To assess whether this is feasible, we evolve the system using the equilibrium tide framework, Eqs.~2--12 from \citet[][]{2010A&A...516A..64L} for a range of initial planetary obliquities, $\epsilon_{p, i}$ and $Q'_p$ values (see Appendix~\ref{app:tides} for details). For each $(Q'_p, \epsilon_{p,i})$ pair, we track the $\epsilon_p$ evolution and record the damping timescale $\tau_{\rm damp}$. We define $\tau_{\rm damp}$ as the time at which the planetary obliquity is fully damped, i.e., $\epsilon_p\rightarrow0^\circ$. 

We find that the region of parameter space enclosed by the cyan contour in Figure~\ref{fig:Qp_vsep_meshplot}, where $L_{\rm tide} \gtrsim L_{\rm tide, req}$, undergoes very rapid obliquity damping, with ${\tau_{\rm damp} \lesssim 10^4}$~yr. Even if we assume an age range (10-30~Myr) for V1298 Tau, and typical disk dissipation timescales of $\sim$10~Myr \citep{mamajek2009}, it is unlikely that the planet could retain this obliquity given the decay timescale is shorter than 1Myr by a factor of 100.



\subsubsection{Spin--orbit resonant locking}
Although tides damp obliquity, planets can sometimes maintain non-zero values if they are trapped in a secular spin–orbit resonance \citep{2019NatAs...3..424M, 2022MNRAS.509.3301S}. This mechanism has been invoked to explain obliquity excitation for several Solar System planets \citep{1993Natur.361..608L, 1993Sci...259.1294T, 2004AJ....128.2501W, 2006ApJ...640L..91W} and has recently been studied in short-period exoplanets \citep[e.g.][]{2019NatAs...3..424M, 2020ApJ...903....7S, 2020ApJ...905...71M, 2022MNRAS.509.3301S, 2022MNRAS.513.3302S, 2024ApJ...975..256G, 2024ApJ...961..203M}. We therefore investigate whether V1298 Tau b could be trapped in such a resonance, allowing it to sustain a high obliquity over timescales longer than the nominal tidal damping time. 

To assess this possibility, we compare the planet’s spin precession frequency, set by the torque exerted by the host star on the planet’s rotational bulge, with the orbital nodal precession frequency modes, $g = \dot{\Omega}$, where $\Omega$ is the longitude of the ascending node. The nodal precession can be driven by several mechanisms; here, we focus on planet–planet interactions given that V1298 Tau is a multi-planets system. Resonant configurations, known as ``Cassini states", can arise when the spin and orbital precession frequencies become commensurate, allowing the spin axis to maintain a fixed obliquity despite ongoing dissipative processes.

The relevant quantity for the planetary spin precession is the precession constant $\alpha$, given by \citep{1997A&A...318..975N, 2003Icar..163....1C}
\begin{equation} \label{eq:alpha}
    \alpha = \frac{1}{2}\frac{M_\star}{M_p}\left(\frac{R_p}{a} \right)^3\frac{k_2}{C} \frac{\omega_p}{(1 - e^2)^{(3/2)}}.
\end{equation}
Here, $M_\star$ is the stellar mass, $a$ is the semi-major axis, $M_p$ and $k_2$ are the planet's mass and Love number respectively, and $C$ is the planet’s dimensionless moment of inertia, normalized by $M_pR_p^2$. In evaluating $\alpha$, we set the planetary spin rate to its equilibrium value, $\omega_p=\omega_{\rm eq}$, given by \citep{2010A&A...516A..64L}:
\begin{equation} \label{eq:omega_eq}
    \omega_{\rm eq} = n \frac{N(e)}{\Omega(e)}\frac{2 \cos{\epsilon_p}}{1 + \cos^2{\epsilon_p}}.
\end{equation}
where $n$ is the planet's mean motion, and functions--$N(e), \Omega(e)$ are defined in Appendix~\ref{app:tides}. 

The $k_2$ for V1298 Tau b is currently unconstrained. We therefore adopt a range of fiducial values, $k_2 \in [0.1, 0.5]$, motivated by estimates for Solar System planets, e.g. Saturn ($k_2 = 0.39$), Uranus ($k_2 = 0.104$), and Neptune ($k_2 = 0.127$) \citep{2016CeMDA.126..145L}. For each choice of $k_2$, we compute $C$, using the Radau–Darwin relation, which relates the moment of inertia coefficient to the planet’s first-order tidal distortion and is given by \citep{2011A&A...528A..18K}:
\begin{equation}
    C \approx \frac{2}{3}\left[1 - \frac{2}{5}\left(\frac{5}{k_2+1} - 1 \right)^{(1/2)} \right].
\end{equation}

Next, we compute the nodal precession frequency modes using the secular Laplace-Lagrange approximation \citep{1999ssd..book.....M}. Since the V1298 Tau system contains four observed planets, the nodal precession is decomposed into three eigenfrequencies, $g_i$ with $i \in \{ 1, 2, 3\}$. In the limit of small eccentricities and inclinations, and for planets not in mean-motion resonances, $\Omega$ evolves as a superposition of these secular eigenmodes. In Laplace-Lagrange theory, $g_i$ are obtained as the eigenvalues of the nodal matrix \citep{1999ssd..book.....M} (see Appendix~\ref{app:LL} for details of matrix construction). For the V1298 Tau system, we obtain $g_{1,2,3} = \{0.0163, 0.0056, 0.0024\}$~${\rm yr}^{-1}$.

\begin{figure}
    \centering
    \includegraphics[width=\linewidth]{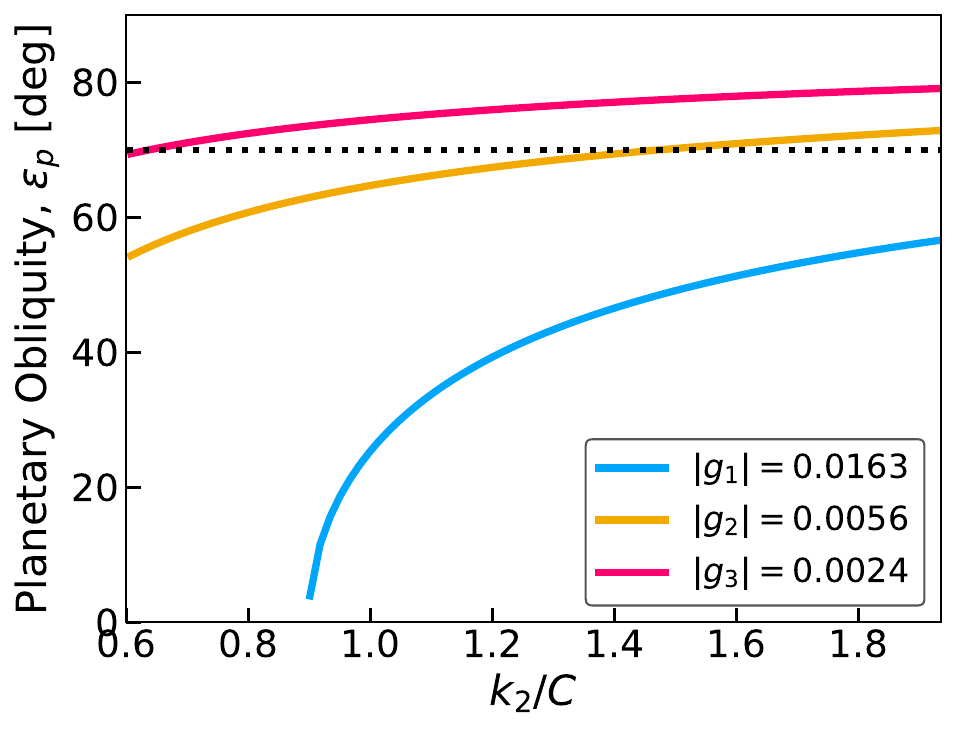}
    \caption{Obliquity of Cassini state 2 vs. $k_2/C$ for each nodal precession frequency $g_i$ (color-coded). The black dotted line marks $\epsilon = 70^\circ$, above which tidal heating can account for the observed internal temperature (at $Q'_p \sim 10^2$).}
    \label{fig:cassini state obl}
\end{figure}

To identify possible spin-orbit resonances, we solve the resonance condition \citep{2024ApJ...961..203M}:
\begin{equation} \label{eq:resonance_locking}
    |g_i| \approx \alpha\cos{\epsilon_p}.
\end{equation}
For each nodal frequency mode $g_i$, the resonant obliquity depends on $\alpha$, which in turn depends on $\epsilon_p$ because we evaluate $\alpha$ using $\omega_p=\omega_{\rm eq}(\epsilon_p)$. Therefore, the resonance condition must be solved self-consistently for $\epsilon_p$. Combining Eq.~\ref{eq:omega_eq}, and Eq.~\ref{eq:resonance_locking} gives:
\begin{equation} \label{eq:cassini_state_obliquity}
    \cos{\epsilon_p} = \left[ 2\frac{N(e)}{\Omega(e)}\frac{\alpha_{\rm syn}}{|g_i|} - 1 \right]^{-1/2},
\end{equation}
where $\alpha_{\rm syn} = \alpha(n/\omega_p)$. For a given $g_i$ and fixed value of $k_2/C$, Eq.~\ref{eq:cassini_state_obliquity} gives the stable equilibrium resonant obliquity in Cassini state 2. Since $\alpha\propto k_2/C$, evaluating this expression for a range of plausible $k_2/C$ values produces a resonance curve in the $k_2/C-\epsilon_p$ plane for each eigenmode, as shown in Figure~\ref{fig:cassini state obl}. These curves illustrate the range of obliquities at which the planet could plausibly be resonantly locked. 

We showed earlier that explaining the observed $T_{\rm int}$ requires $\epsilon_p\gtrsim70^\circ$ even for $Q'_p\sim10^2$; we thus focus only on resonant configurations above this threshold. The $g_3=0.0024~{\rm yr}^{-1}$ mode permits such high-obliquity locking over the full adopted range of $k_2/C$, and the $g_2=0.0056~{\rm yr}^{-1}$ mode permits it for $k_2/C\gtrsim1.5$, corresponding to $k_2\gtrsim0.33$. Although V1298 Tau b's spin axis could in principle be locked in other modes, those configurations occur at lower obliquities and therefore cannot provide the required tidal luminosity. 

While at first glance these high-obliquity configurations seem like plausible sources of the high internal temperature, their existence is not guaranteed because high-obliquity resonances cannot be sustained for arbitrarily strong tidal dissipation. Tidal damping disrupts the Cassini state and breaks the resonance when $Q_p \lesssim Q_{\rm break}$, where $Q_{\rm break}$ is a critical tidal quality factor given by \citep{2024ApJ...961..203M}
\begin{equation}
    Q_{\rm break} \approx \frac{3}{4}\frac{\sin{\epsilon}(1 + \cos^2{\epsilon})}{\cos^3{\epsilon}}.
\end{equation} 
We convert $Q'_p$ to the tidal quality factor, $Q_p=2k_2Q'_p/3$. Keeping $Q'_p=10^2$ fixed, we evaluate $Q_p$ over the adopted range of $k_2$ and compare it to $Q_{\rm break}$ for each resonant solution.


For the $g_3$ mode, tidal dissipation is strong enough to break the resonance across the full range of $k_2/C$ values considered. Thus, although the $g_3$ mode can reach $\epsilon_p\gtrsim70^\circ$, it is unlikely to sustain the resonance when $Q'_p\sim10^2$. In contrast, the $g_2$ mode remains stable against tidal disruption over the relevant high-obliquity portion of parameter space. Therefore, V1298 Tau b can maintain $\epsilon_p\gtrsim70^\circ$ and produce $L_{\rm tide}\gtrsim L_{\rm tide,req}$ only if it is locked to the $g_2$ mode with $k_2/C\gtrsim1.5$ (or $k_2\gtrsim0.33$), and if $Q'_p\sim10^2$.

We note, however, that such a low $Q'_p$ is extreme and likely not representative of V1298 Tau b's properties. Improved constraints on tidal dissipation in young planets are essential to determine whether such low $Q'_p$ values are physically plausible. Thus, tidal heating remains a possible explanation for the inferred high $T_{\rm int}$, but only under finely tuned conditions, and is unlikely to explain the growing sample of planets requiring elevated intrinsic luminosity.

\subsection{Evolution models with composition gradient for V1298 Tau b}

\subsubsection{Overall concept} \label{concept}

Thermal evolution models have assumed adiabatic deep interior temperature-pressure profile, where heat is transported from the interior to the radiative layers through efficient convection \citep{Guillot2005,guillot2010,muller2026}. The assumption of an adiabatic interior thermal profile is consistent with a core-envelope interior structure (heavy metals in the core with an overlying H/He dominated envelope). Atmospheric forward models for V1298 Tau b (and in general) assume an adiabatic T-P profile, which allows us to infer intrinsic temperature from the temperature at the quench pressure. However, if there are metallicity (mean molecular weight) gradients in the interior, the thermal profile could become non-adiabatic due to inefficient convection \citep{ledoux1947,leconte2017,misener2022}. As a result, heat transport to the upper atmosphere is weak and leads to a prolonged cooling timescale for the planet \citep{chabrier2007,vazan2016}.

Metallicity gradients in the interior could be a natural outcome of planet formation through pebble accretion \citep{ormel2021,bloot2023}. One of the potential sources for metallicity gradients is the rainout of silicates. The presence of silicate vapor in the interior could drive the thermal profile to a non-adiabatic regime \citep{misener2022,Markham2022}. Silicate particles which are expected to be present in a young planetary envelope after formation could condense (rainout) over Gyr timescales \citep{vazan2024} which result in a metallicity gradient and deposit heat in the interior due to the release of gravitational potential energy and latent heat of condensation of silicate vapor \citep{vazan2023, vazan2024}. Similarly, helium rain, as hypothesized for Saturn could also significantly affect the deep thermal profile \citep{howard2024}. Deep metallicity gradients have been suggested for solar system gas giants \citep{wahl2017,mankovich2021,morf2025}. The low luminosity of Uranus has also been explained by considering metallicity gradients in its envelope \citep{vazan2020}. Therefore, the existence of metallicity gradients is expected for exoplanets as well, particularly when they are young.

In the following sections we calculate and compare the thermal and structural evolution of V1298 Tau b assuming three different interior models. First, we compute a standard core-envelope model, where the metals are concentrated within the `core' with an overlaying H/He envelope. Such a structure leads to an adiabatic T-P profile. Secondly, keeping the core-envelope model as a reference, we compute a model containing a stabilizing heavy-element gradient. In regions that are stable to large-scale overturning convection according to the Ledoux criterion but unstable according to the Schwarzschild criterion \citep{kippenhahn2013}, we assume that heat is transported by layered oscillatory double-diffusive convection (DDC; \citealt{chabrier2007,Rosenblum2011,leconte2013,leconte2017}).
 Lastly, we compute a model with silicate rainout which can affect the T-P profile both by inhibiting deep convection as well as deposit energy due to gravitational settling.

For each model, the initial conditions are set by inputs from planet formation models \citep{ormel2021}. We assume a mass of 12~M$_{\oplus}$, 10$\times$solar atmospheric metallicity, which sets the opacity at the current age of the planet. We discuss the model calculations in the following subsection.


\subsubsection{Interior structure and evolution models} \label{structure models}

To model the interior structure and its evolution we use the formalism as outlined in \citet{vazan2024}. The model couples a planetary thermal evolution code \citep{vazan2015} with an interior structure model for planets with metal-rich interiors \citep{vazan2018}. Our model allows for heat transport by convection, radiation and conduction, depending on the local conditions in time. The thermal evolution is calculated for the entire
interior (from the center to surface) on one mass grid, with no assumption of ``core" and ``envelope" regions. The interior structure can evolve (redistribution of composition) by convective-mixing, when convection criterion is fulfilled \citep{vazan2015}, or by silicate rainout in over-saturated envelope layers \citep{vazan2024}. The initial models are derived from planet formation calculations for sub-Neptune planets that formed with composition gradients in their envelopes due to pebble accretion \citep{ormel2021}. To explain the current mass (12~M$_{\oplus}$) and radius (9.41~R$_{\oplus}$) we require 35\% by mass H/He envelope, consistent with mass-radius scaling relations \citep{Lopez_2014}. Atmospheric mass loss due to photoevaporation is included using the formalism outlined in \citet{Rogers2021} assuming energy-limited photoevaporation. We assume energy-limited photoevaporation, with a mass-loss efficiency of 0.1. We assume an atmospheric metallicity of 10 $\times$ solar \citep{barat2025}. This metallicity of 10$\times$Solar is used to set the mean opacity of the atmosphere assuming the formalism outlined in \citet{Freedman2014}. The models are initialized such that they match the current mass-radius of the planet.

\subsubsection{Comparing thermal profiles between different interior structures} \label{structure_tp_profiles}

\begin{figure*}
    \centering
    \includegraphics[width=1\textwidth]{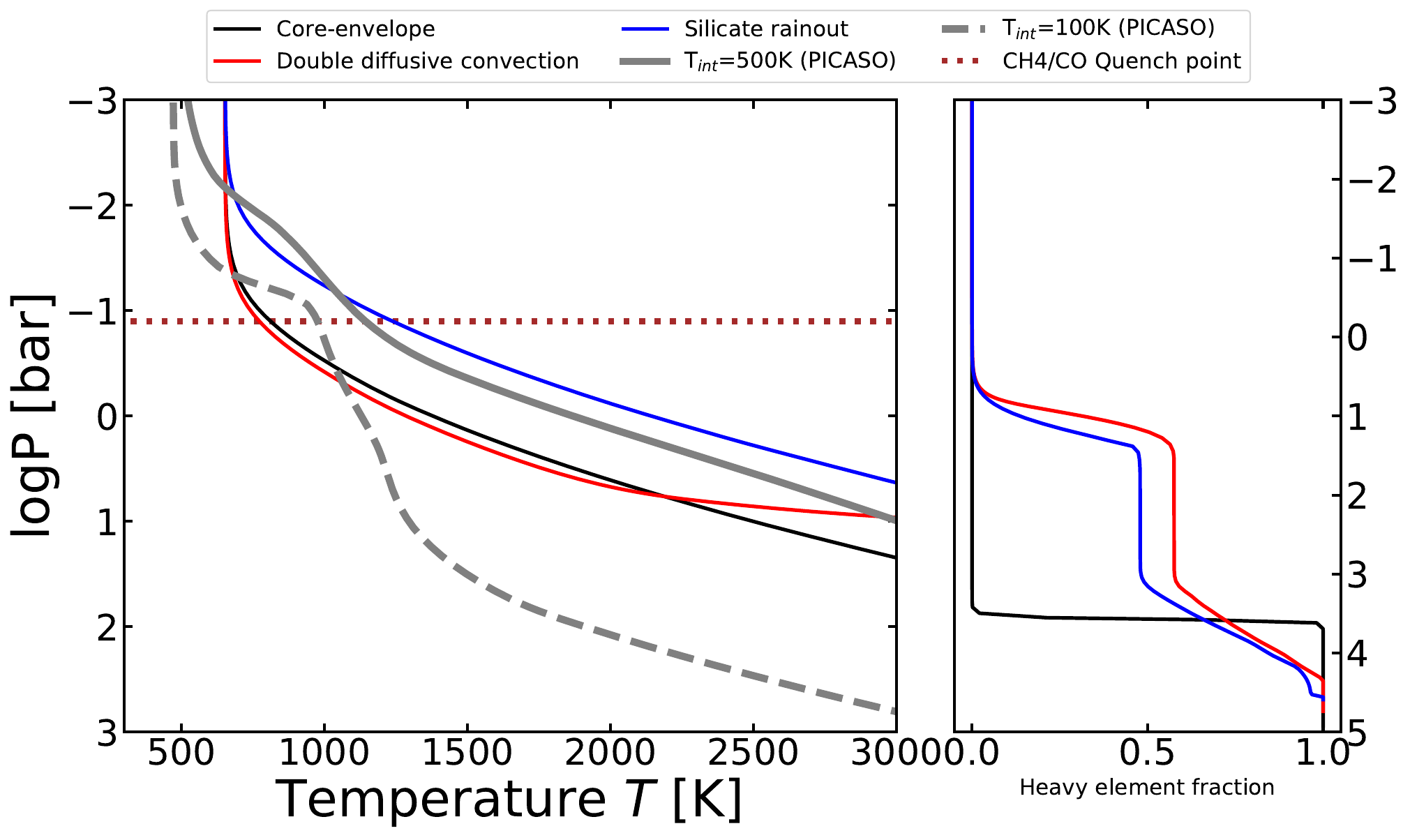}
    \caption{We compare the T-P profiles (left) for V1298 Tau b simulated using metallicity gradient evolution models described in Section \ref{structure_tp_profiles} and corresponding mass distribution (normalized) within the envelope is shown in the right panel. The simulations are outlined in Section \ref{structure models}. The black, red and blue models show T-P profiles for the core-envelope, double diffusive convection and silicate rainout models at 23~Myr (current age of system). The grey lines show self-consistent forward models generated using the radiative-convective equilibrium model \texttt{PICASO} \citep{Mukherjee22} from \citet{barat2025}. The solid line shows the best-fit T$_{int}$ model (500~K) and the dashed line shows a model for 100~K. At the quench pressure for CH$_4$ \citep{barat2025} the T$_{int}$=500~K is hotter than the 100~K model by $\sim$ 300~K. At this pressure the silicate rainout model shows a similar deviation from the adiabtic core-envelope model, but the DDC remains similar to the adiabatic model. }
    \label{fig:T-P_comparison}
\end{figure*}

We compare the T-P profiles from three models with different interior structures --- core-envelope (CE), double diffusive convection (DDC) and silicate rainout (SR) at the current age of V1298 Tau b in Figure~\ref{fig:T-P_comparison}. Metallicity gradients in the atmosphere could inhibit large-scale convection and lead to small-scale layered/double diffusive convection (DDC, \citealt{Rosenblum2011,leconte2012}). We also show self-consistent T-P profiles for different internal temperatures (100~K and 500~K) from PICASO \citep{batalha19} for V1298 Tau b from \citet{barat2025}. The quench point pressure derived from fitting its transmission spectrum \citep{barat2025} using this grid is $\sim$0.1 bar with an intrinsic temperature of 500~K. The PICASO grid assumes an adiabatic (convective interior) T-P profile which allows us to constrain the intrinsic temperature from the temperature at the quench pressure for CH$_4$. This assumption is only consistent with the core-envelope assumption. 


We find the T-P profile for the DDC model deviates from the adiabatic CE model and becomes super-adiabatic at pressures higher than 10 bar.  To remove methane from the upper atmosphere for the CE and DDC T-P profiles, the quench pressure is likely to be an order of magnitude deeper than the 0.1~bar. Such a deep quench point would require much higher vertical mixing, and would not be able to match observed abundances of other molecules.  Therefore, we can conclude that DDC and CE are less likely to reproduce the high temperature needed at lower observable pressures to explain the methane abundance of V1298 Tau b, without additional energy sources.

On the other hand, models with silicate rainout can affect the T-P profile at much lower pressures compared to the DDC model (Figure \ref{fig:T-P_comparison}). In this model the T-P profile is affected due to the release of gravitational potential energy of settling silicate grains in the young envelope. We note that this model also has an initial metallicity gradient from planet formation models. Therefore, the added effect of silicate rain results in the T-P profile deviating from the adiabatic profile at relatively low pressures. At the quench pressure for V1298 Tau b, the silicate rainout T-P profile is hotter than the adiabatic T-P profile by $\sim$700~K, and is sufficient to flip the carbon chemistry from CH$_4$ dominated to CO/CO$_2$ dominated \citep{mukherjee2024}. Using scaling relations presented in \citet{vazan2024} we estimate an approximate silicate rainout luminosity of 10$^{25}$~ergs/s assuming a gas-to-core ratio of 1\% \citep{barat2025}. This is lower than the luminosity estimated for tidal heating, but silicate rainout energy deposition happens high up in the envelope, and the presence of metallicity gradients can trap this energy to increase the local temperature. Therefore, we conclude that including silicate rainout in an envelope with metallicity gradients, could potentially explain the high temperature of V1298 Tau b in the deep atmosphere without requiring additional heating sources. 

\begin{figure}
    \centering
    \includegraphics[width=1\linewidth]{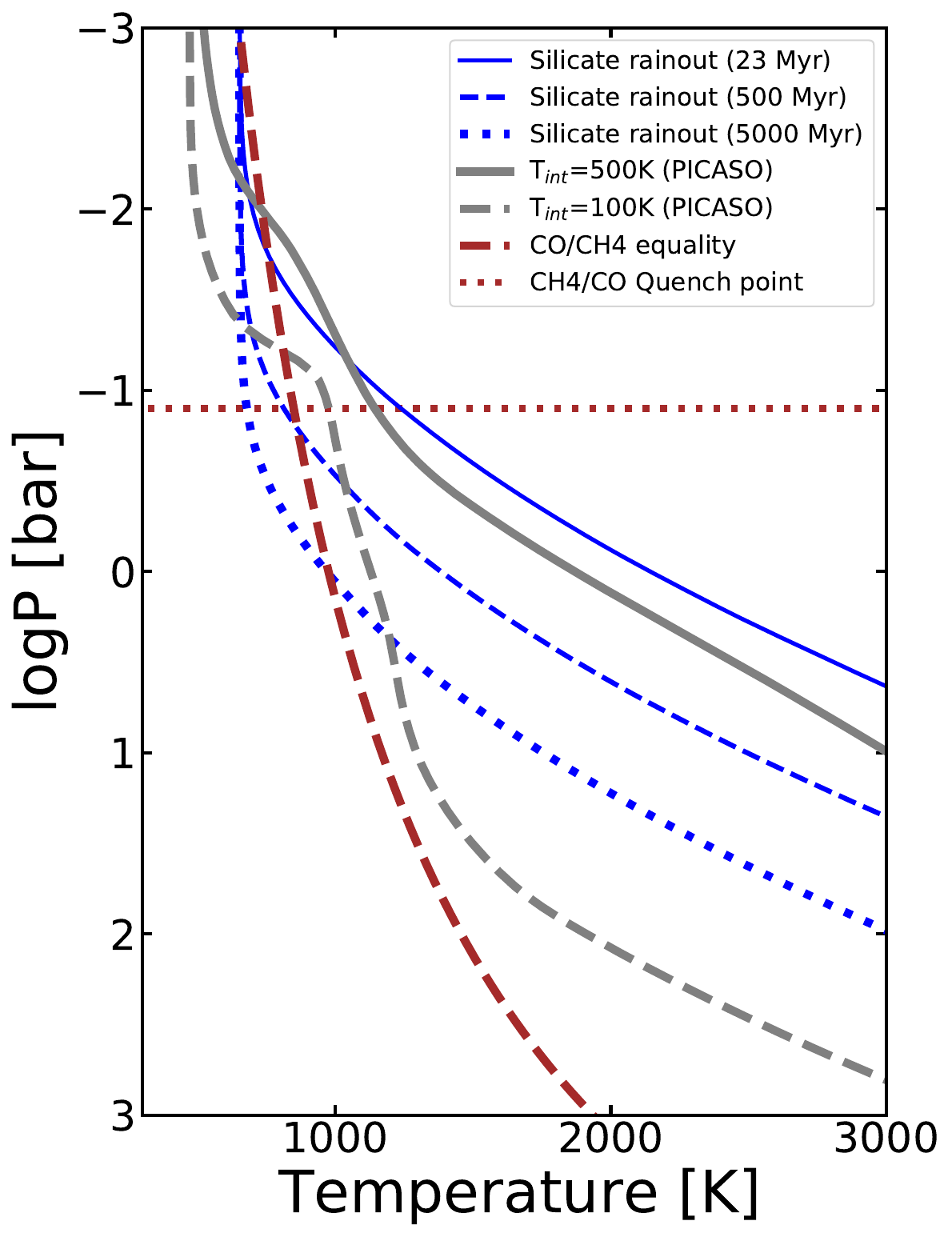}
    \caption{We show the simulated T-P profiles for V1298 Tau b assuming silicate rainout (Section \ref{structure_tp_profiles}) at three points in its evolution --- 23~Myr (current age), 500~Myr and 5 Gyr. We show the PICASO T$_{int}$=100 and 500~K for comparison. The brown dotted line shows the inferred methane quench point from \citet{barat2025}. The brown dashed line shows the CO/CH$_4$ equality line adapted from \citet{fortney_2020}. These models show that within 500~Myr the silicate rainout T-P profile cools down significantly and the deep atmosphere (0.1-1 bar) could go from CO dominated ton CH$_4$ dominated regime.}
    \label{fig:t-p_evolution_v1298}
\end{figure}

We model the evolution of V1298 Tau b including silicate rainout. The silicate rainout timescale is primarily driven the envelope mass fraction and to a lesser extent by planet mass and atmospheric metallicity \citep{vazan2024}. Our JWST observations constrain these parameters for V1298 Tau b, and our evolutionary model is benchmarked at these values. We compare the T-P profiles at the current age (10-30 Myr), 500~ Myr and 5000~Myr in Figure \ref{fig:t-p_evolution_v1298}. Our simulation shows that within 500~Myr the interior would undergo significant cooling. At this age ($>$500~Myr) it would be difficult to remove sufficient methane due to vertical mixing. Therefore, if the current lack of methane is due to silicate rainout, the methane feature could emerge in this planet (assuming the quench pressure remains the same) as it evolves.

\subsection{Effect of cloud condensation of deep T-P profile of V1298 Tau b} \label{cloud condensation}

Presence of optically thick clouds at deep pressures have been hypothesized for sub-Neptune exoplanets \citep{lee2025}. Recent self-consistent atmosphere and interior structure models, that include clouds and their radiative feedback, have shown that temperatures in the deep atmosphere of sub-Neptunes ($>$100 bar) could be higher by more than 1000~K compared to clear atmosphere models \citep{mukherjee2026}. These clouds were also shown to have a substantial impact on the phase (solid vs. magma oceans) and the chemical composition of the atmosphere-mantle interface for sub-Neptunes \citep{mukherjee2026}. Deep optically thick clouds trap the thermal radiation from the planet's interior, heating up the deep atmosphere and moving the radiative-convective boundary to much smaller pressures relative to clear models. \citet{mukherjee2026} has proposed that this effect will likely bias the intrinsic temperature measured with clear atmosphere models \citep[e.g.,][]{barat2025} by increasing the temperature at the CH$_4$/CO quench point. 

\begin{figure*}
    \centering
    \includegraphics[width=1\textwidth]{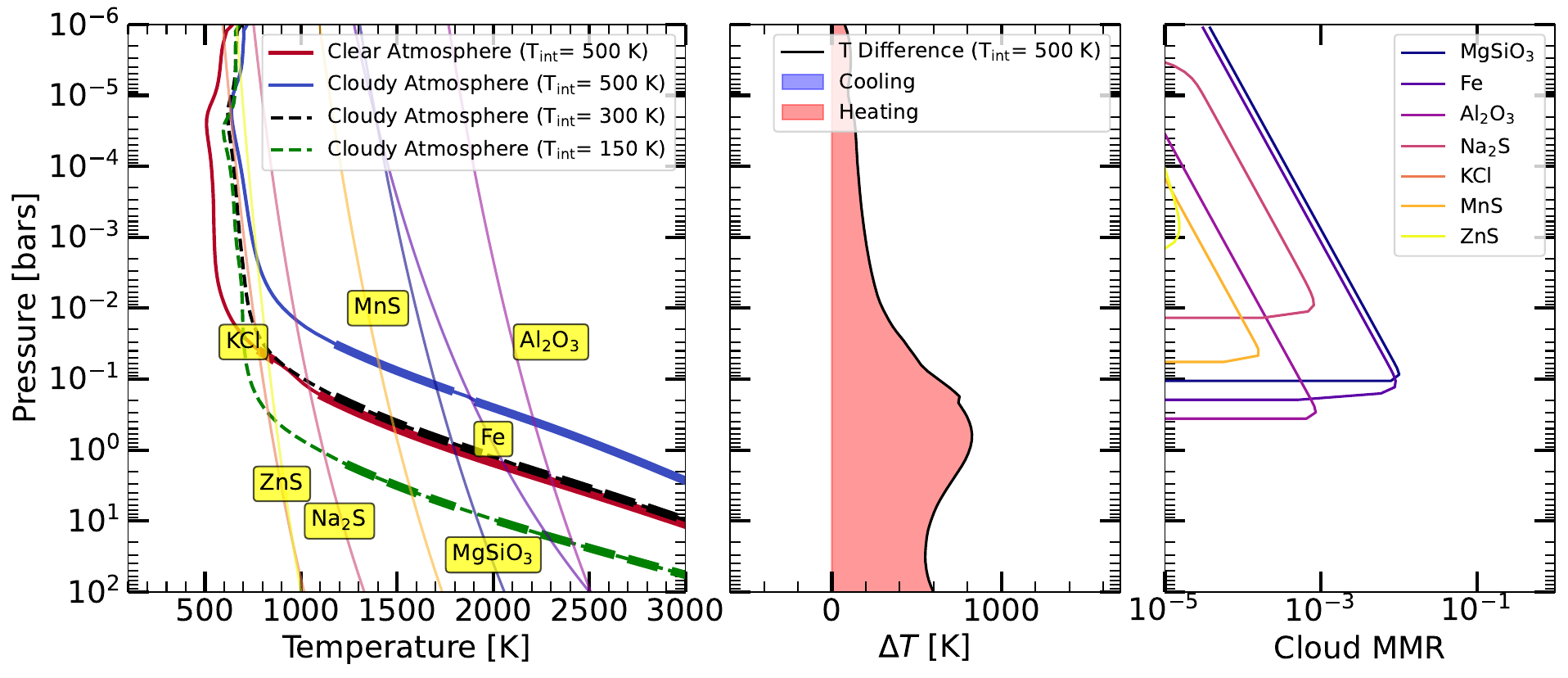}
    \caption{Self-consistent T-P profiles for V1298 Tau b generated using the \texttt{PICASO} framework for T$_{int}$=150~K (green dashed), 300~K (black dashed) and 500~K. We include cloud condensation for all three models and also show the clear 500~K intrinsic temperature model (red solid). We find that incorporating cloud condensation pushes the T-P profile to higher temperatures for a given intrinsic temperature. A clear 500~K model and 300~K model with cloud condensation appear similar for pressures deeper than 0.1~bar. The different color lines show the condensation curve for different mineral species. The middle panel shows the difference between a 500~K intrinsic temperature model with and without clouds. The right panel shows the mass mixing ratio of the condensate species considered in our model. }
    \label{fig:cloud_condensation}
\end{figure*}

We compute cloudy radiative-convective atmosphere models using \texttt{PICASO} \citep{Mukherjee22,mang2026}, asuming 10$\times$solar metallicity, 0.3 C/O ratio and stellar SED from \citet{Duvvuri2023} at three intrinsic temperatures --- 150~K, 300~K and 500~K shown in Figure \ref{fig:cloud_condensation}. We include clouds composed of MgSiO$_3$, Fe, Al$_2$O$_3$, Na$_2$S, KCl, MnS, and ZnS and assume 10$\times$solar metallicity. For 150~K and 300~K, we only compute models which include deep cloud condensation. We compute both cloudy and clear models for $T_{\rm int}= 500$~K. We note that we assume a spatially uniform cloud cover in these models, ignoring high levels of spatial inhomogeneities in cloud-coverage recently observed in several hot Jupiters \citep[e.g.,][]{Mukherjee2026limb,Fu2025}. 

We find that the T-P profile for a clear T$_{\rm int}$=500~K model atmosphere matches very well with a T$_{\rm int}$=300~K cloudy model for pressures deeper than 0.1~bar (Figure \ref{fig:cloud_condensation}). The models generated in \citet{mukherjee2026} were tailored for high metallicity sub-Neptunes. Here we test whether this same mechanism can explain the high intrinsic temperature measured for V1298 Tau b --- a H/He dominated low metallicity atmosphere unlike the mature sub-Neptunes.

The cloud radiative feedback in a model atmosphere with an intrinsic temperature of 150~K is not sufficient to heat the T-P profile to sufficiently high temperatures near the quench pressure ($\sim$0.1 bar) to match the 500~K clear atmosphere model. At V1298 Tau b's current age, the expected intrinsic temperature from a core-envelope structure (Figure \ref{fig:figure1}) is $\sim$150~K. Therefore, deep cloud condensation can reduce the discrepancy between predicted (150~K) and inferred (500~K) intrinsic temperatures at $\sim$10$\times$solar metallicity, but we still require additional heating/heat trapping mechanisms to account for the required high temperature at the quench point of methane.




\subsection{Potential explanation of high intrinsic temperature of WASP-107b}

WASP-107b, a mature super-Neptune (Table \ref{tab:table1}) has a transmission spectrum very similar to V1298 Tau b (Figure \ref{fig:figure1}), and can similarly be explained with a high inferred intrinsic temperature (450~K) \citep{welbanks2024,sing2024}. In this section we evaluate the three hypotheses (tidal heating, non-adiabatic T-P profiles due to metallicity gradient and deep cloud condensation) in the context of WASP-107b. Recent RV follow-up constrains its eccentricity to an upper limit of 3\% \citep{yee2026} compared to $\sim$13\% \citep{piaulet2021}. At 3.4~Gyr it is unlikely to have significant spin obliquity. Therefore, we do not consider tidal heating for WASP-107b.

\begin{figure}
    \centering
    \includegraphics[width=1\linewidth]{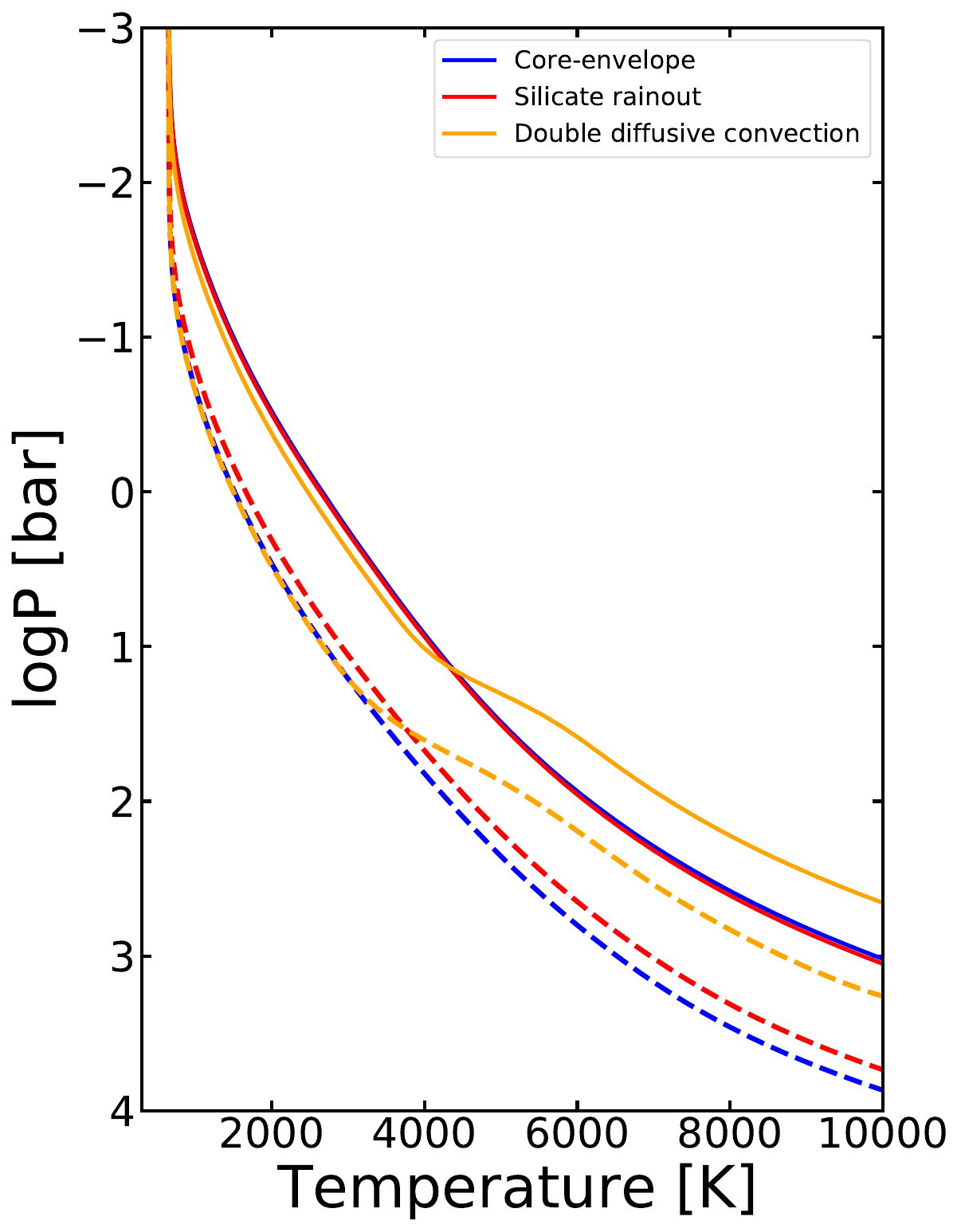}
    \caption{We show the simulated T-P profiles for WASP-107b for three internal structure models: core-envelope, silicate rainout and double diffusive convection. We find that silicate rainout does not have a significant effect on the T-P profile of this planet which is more massive compared to V1298 Tau b. While DDC has an effect it appears at pressures deeper than 10 bar and is unlikely to affect the chemistry in the upper atmosphere. The solid lines are at 20~Myr and the dashed lines at the current age (3 Gyr). The T-P profiles for WASP-107b cool down significantly at its current age, unlike V1298 Tau b. }
    \label{fig:t-p_wasp107b}
\end{figure}

In Figure \ref{fig:t-p_wasp107b} we show initial and current T-P profiles simulated for WASP-107b assuming core-envelope structure, silicate rainout and double-diffusive convection. We find that silicate rainout does not have a significant effect on the T-P profile of WASP-107b at any age. This is most likely due to the much higher mass of WASP-107b ($\sim$30~M$_{\oplus}$), compared to V1298 Tau b (12~M$_{\oplus}$) --- a higher mass envelope requires higher amount of energy release from settling silicates. Therefore, unlike V1298 Tau b, silicate rainout is not expected to have a significant effect on the T-P profile of WASP-107b, even at very young age. Models including double-diffusive convection deviate from the adiabatic core-envelope T-P for pressures deeper than 10~bar for both initial and current age models. However, the quench pressure is higher up in the atmosphere, and there is unlikely to affect the chemical composition of the observable upper atmosphere. 

On the other hand deep cloud condensation is expected to operate similarly in WASP-107b as shown for V1298 Tau b as both planets have similar temperature and metallicity. Therefore, it could reduce the discrepancy between inferred and measured T$_{int}$ for this planet. 




\section{Discussion} \label{discussion}

\subsection{Emerging trend of elevated inferred intrinsic temperature}

We find that both silicate rainout and deep cloud condensation have a similar effect --- they can increase the temperature in the deep atmosphere without a necessity of increasing the intrinsic luminosity of the planet. An important point of distinction between these two mechanism is the age dependence. While the effect of silicate rainout is expected to decrease with age, heat trapping due to deep cloud condensation can occur at all ages. Observations have revealed high inferred intrinsic temperature both at young age for V1298 Tau b and e \citep{barat2025,dai2026} as well as mature planets such as WASP-107b \citep{sing2024,welbanks2024} and GJ3470b \citep{beaty2024}. In fact, elevated inferred intrinsic temperatures could be relatively common for the exoplanet population \citep{Baxter2021,yu2026}.   

Our models show that cloud condensation alone is not sufficient to explain the high inferred intrinsic temperature of V1298 Tau b. On the other hand, evolutionary models such as silicate rainout are unlikely to have a significant impact for mature planets like WASP-107b. Therefore, it is possible that in reality both these mechanisms --- settling of silicates/helium releasing latent heat and heat trapping due to cloud condensation in the deep atmosphere --- may occur in concert, but dominate the deep atmosphere T-P profile at different stages of evolution. In this case we expect the inferred intrinsic temperature to evolve with age for the first few hundred million years, and `plateau' after that. The transition from silicate rainout to atmospheric chemistry driven processes like deep cloud condensation could occur around the timescale for gravitational settling of silicates in the atmosphere. It is important to map the evolution of intrinsic temperature, as this would lead to slower contraction for the planet, which could lead to stronger atmospheric mass loss. Future atmospheric characterization of warm transiting planets between 20-500~Myr could help us to constrain the evolution of the inferred intrinsic temperature.


\subsection{Measuring the internal flux of V1298 Tau b}

The intrinsic temperature inference of V1298 Tau b is achieved through the `methane thermometer' \citep{fortney_2020}, which is an indirect approach. The retrieved isothermal temperature of V1298 Tau b from atmospheric retrievals \citep{barat2025} is $\sim$500~K, comparable to the required intrinsic temperature ($\sim$500~K). The emergent flux from the planet is given  by:

\begin{equation*} \label{eq:eq1}
    F_{p} = \sigma~(T_{iso}^{4}+T_{int}^{4})
\end{equation*}

where T$_{iso}$ is the isothermal temperature of the planet and T$_{int}$ is the intrinsic temperature. Normally, the intrinsic temperature is much lower than the isothermal temperature and the thermal emission from the planet is dominated by the isothermal temperature. But, for V1298 Tau b both these temperatures are comparable. Therefore, the emergent flux from the planet during secondary eclipse is expected to be sensitive to the intrinsic temperature.

\begin{figure*}
    \centering
    \includegraphics[width=1\textwidth]{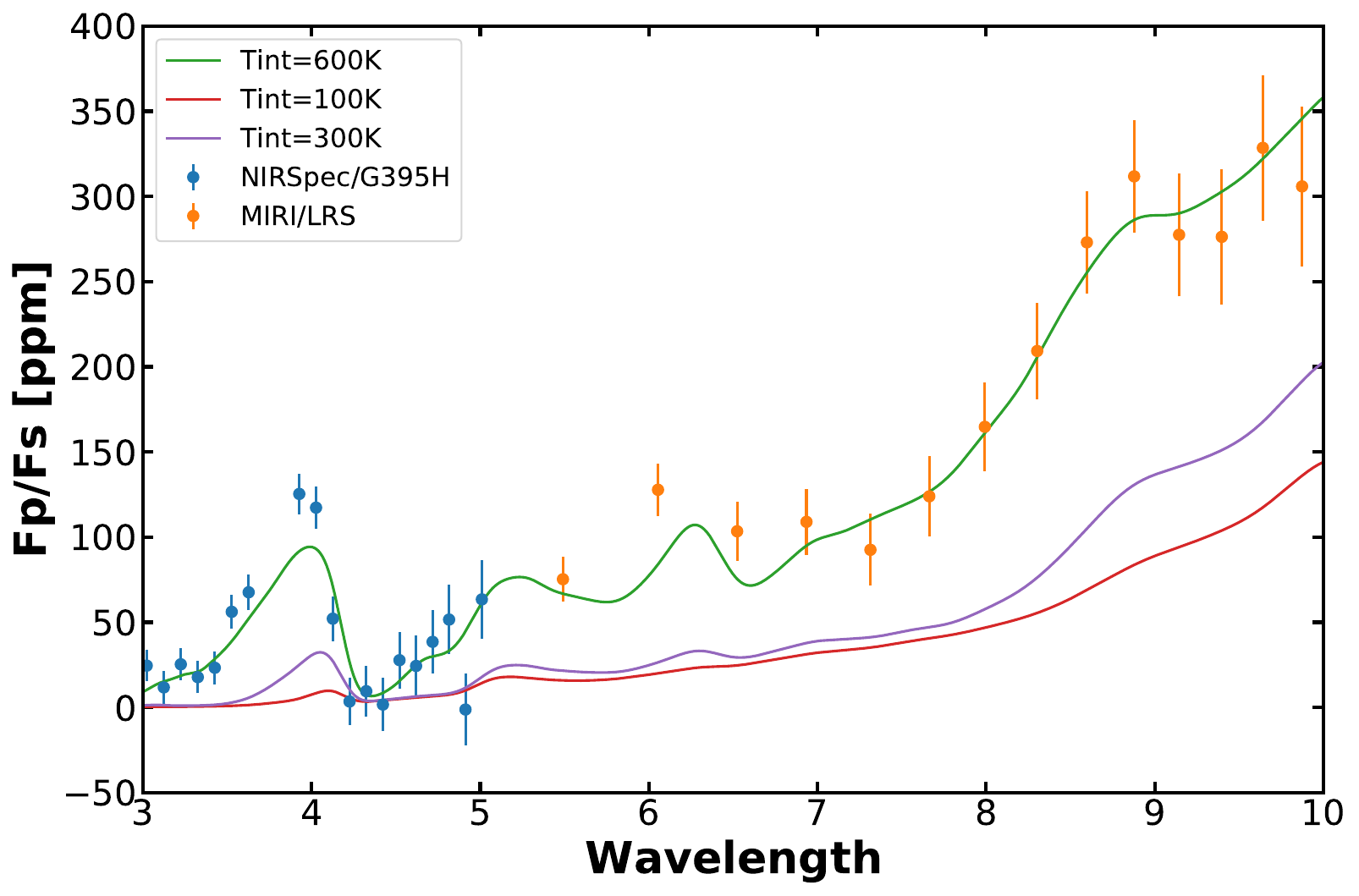}
    \caption{Model secondary emission spectra for V1298 Tau b for different intrinsic temperature thermal profiles simulated using PICASO self-consistent forward models. We also show simulated JWST NIRSpec G395H and MIRI/LRS observations for the T$_{int}$=600~K case generated using \texttt{Pandexo} \citep{batalha2019}. }
    \label{fig:figure 2}
\end{figure*}

Therefore, an observation of thermal emission from V1298 Tau b using JWST could directly test whether V1298 Tau b has such a high intrinsic temperature. In Figure \ref{fig:figure 2} we show  simulated thermal emission spectra for V1298 Tau b with NIRSPec/G395H and MIRI/LRS. We calculate these spectra assuming self-consistent T-P profiles from PICASO. We compare the thermal emission spectra for three models having same isothermal temperature ($\sim$500~K) but different intrinsic temperature (100, 300 and 600~K respectively). These three models can be distinguished both in the NIRSpec/G395H bandpass as well as MIRI/LRS. In the NIRSpec bandpass we find a deep absorption feature due to CO$_2$ around 4.3~$\mu$m which has been found in the transmission spectrum \citep{barat2025}. However, the low abundance of methane allows us to peek deep into the atmosphere. Thus, effectively we measure the brightness temperature at two different pressure levels in the NIRSpec bandpass allowing us to measure the T-P profile gradient. While in MIRI/LRS there are no significant opacity sources (due to low methane and SO$_2$ abundance in the transmission spectrum) expected for this planet, which means that we can probe deep atmospheric layers where the absolute flux from different T$_{int}$ is different and is measurable using the precision of MIRI/LRS. The SNR of the mock observations have been simulated using \texttt{Pandexo} \citep{batalha2017}. 


Therefore, if the planet really has a very high intrinsic luminosity (for e.g. due to the existence of spin obliquity), we would detect strong thermal emission in the near and mid-IR. On the other hand, if there are metallicity gradients, the planet has a low luminosity and the thermal emission would be dominated by the T$_{iso}^{4}$ term. These two scenarios can be distinguished using secondary eclipse measurements with JWST NIRSpec and MIRI/LRS.

\section{Summary} \label{conclusion}

\begin{enumerate}
    \item In recent years with JWST observations of transiting planets we are finding growing evidence for high inferred intrinsic temperatures \citep{yu2026} due to a relatively low methane abundance \citep{fortney_2020}, both for young \citep{barat2025,dai2026} and mature planets \citep{sing2024,welbanks2024,beaty2024}. Such high intrinsic temperatures are inconsistent with prediction from core-envelope structure evolution models and require additional heating/heat trapping mechanisms

    \item In this paper we evaluate three hypotheses to reconcile the inferred high intrinsic temperature with the prediction from early evolution models for V1298 Tau b --- tidal heating \citep{millholand_2020}, silicate rainout \citep{vazan2024} and heat trapping due to deep cloud condensation \citep{mukherjee2026}.

    \item Tidal heating can explain the inferred high $T_{\rm int}$ of V1298 Tau b only if several conditions are simultaneously satisfied. In our models, this requires a very low reduced tidal quality factor, $Q'_p\sim100$, and a sustained planetary obliquity, $\epsilon_p\gtrsim70^\circ$. Although such strong tidal dissipation would normally damp the obliquity over timescales much shorter than the system age ($\sim$10$^{4}$ years), the planet could in principle maintain a high-obliquity state if it is locked in a Cassini state. This viable configuration requires resonant locking to a specific nodal frequency mode, with $k_2\gtrsim0.33$. Thus, tidal heating remains physically possible, but only within a finely tuned region of parameter space. Similarly, recent work on WASP-107 b suggests that tides alone may be insufficient to explain its anomalous internal heating, motivating alternative energy sources \citep{batygin2025,yee2026}.

    \item Standard CE model T-P profiles cannot reproduce high temperatures in the deep atmosphere (0.1-1 bar). We find that atmospheric models with a metallicity gradient and heat transport through double diffusive convection also do not show significant deviation at these pressures. For these models to explain the missing methane problem we would require extremely high vertical mixing, which would affect other molecular abundances. It is therefore an unlikely solution.

    \item Young planets could have significant silicates in the atmosphere as a residue of planet formation \citep{ormel2021}; settling silicates could increase the temperature of the deep atmosphere due to conversion of gravitational potential energy to heat as well as heat trapping due to inefficient deep convection. Models with silicate rainout can increase the temperature in the deep atmosphere of V1298 Tau b (0.1-1 bar) by $\sim$700~K compared to adiabatic core-envelope structure T-P profile. This could reconcile the inferred high interinsic temperature of V1298 Tau b with combined planet formation and evolution models. However, silicate rainout is expected to be dominant for younger planets, and is unlikely to explain the observations of WASP-107b and other mature planets.

    \item Deep cloud condensation (0.1-1 bar) could trap heat and increase the thermal evolution timescale. Our models show that including deep cloud condensation reduces the inferred intrinsic temperature for V1298 Tau b to 300~K, which is still higher than the predicted internal temperature assuming a core-envelope scenario (150~K). Therefore, for young planets like V1298 Tau b, deep cloud condensation may reduce the discrepancy between inferred and predicted intrinsic temperature, but still require additional heating/heat trapping mechanisms. However, deep cloud condensation is expected to be relevant for both young and mature planets, and could therefore be a potential explanation for planets like WASP-107b.


\end{enumerate}

\appendix
\section{Tidal Model}
\label{app:tides}
We estimate tidal heating using the equilibrium tide model in the constant time-lag approximation \citep{1980A&A....92..167H}. The tidal luminosity is given by \citet{2010A&A...516A..64L} as follows:
\begin{equation}
    L_{\rm tide}(e,\epsilon) = 2K_p\left[N_{\rm a}(e) - \frac{N^2(e)}{\Omega(e)}\frac{2\rm cos^2\epsilon_p}{1 + \rm cos^2\epsilon_p} \right],
\end{equation}
where $e$ is the eccentricity, $\epsilon_p$ is the planetary obliquity, and $N_{\rm a}(e), N(e)$, and $\Omega(e)$ are defined as
\begin{equation} \label{n_a}
    N_{\rm a}(e) = \frac{1 + \frac{31}{2}e^2 + \frac{255}{8}e^4 + \frac{185}{16}e^6 + \frac{25}{64}e^8}{(1-e^2)^{\frac{15}{2}}}
\end{equation}
\begin{equation} \label{n(e)}
    N(e) = \frac{1 + \frac{15}{2}e^2 + \frac{45}{8}e^4 + \frac{5}{16}e^6}{(1-e^2)^6}
\end{equation}
\begin{equation}
    \Omega(e) = \frac{1 + 3e^2 + \frac{3}{8}e^4}{(1 - e^2)^{\frac{9}{2}}}.
\end{equation}
The quantity $K_p$ is the characteristic luminosity scale given by
\begin{equation} \label{K}
    K = \frac{9n}{4Q_p'}\left(\frac{GM_{\star}^2}{R_p}\right) \left(\frac{R_p}{a}\right)^6,
\end{equation}
where $G$ is the gravitational constant and $n = 2\pi/P_{\rm orb}$ is the mean motion. The reduced tidal quality factor is $Q'_p = 3Q_p/(2k_2)$, where $Q_p$ is the planet's tidal quality factor and $k_2$ is its Love number.

\begin{figure} 
    \centering
    \includegraphics[width=\linewidth]{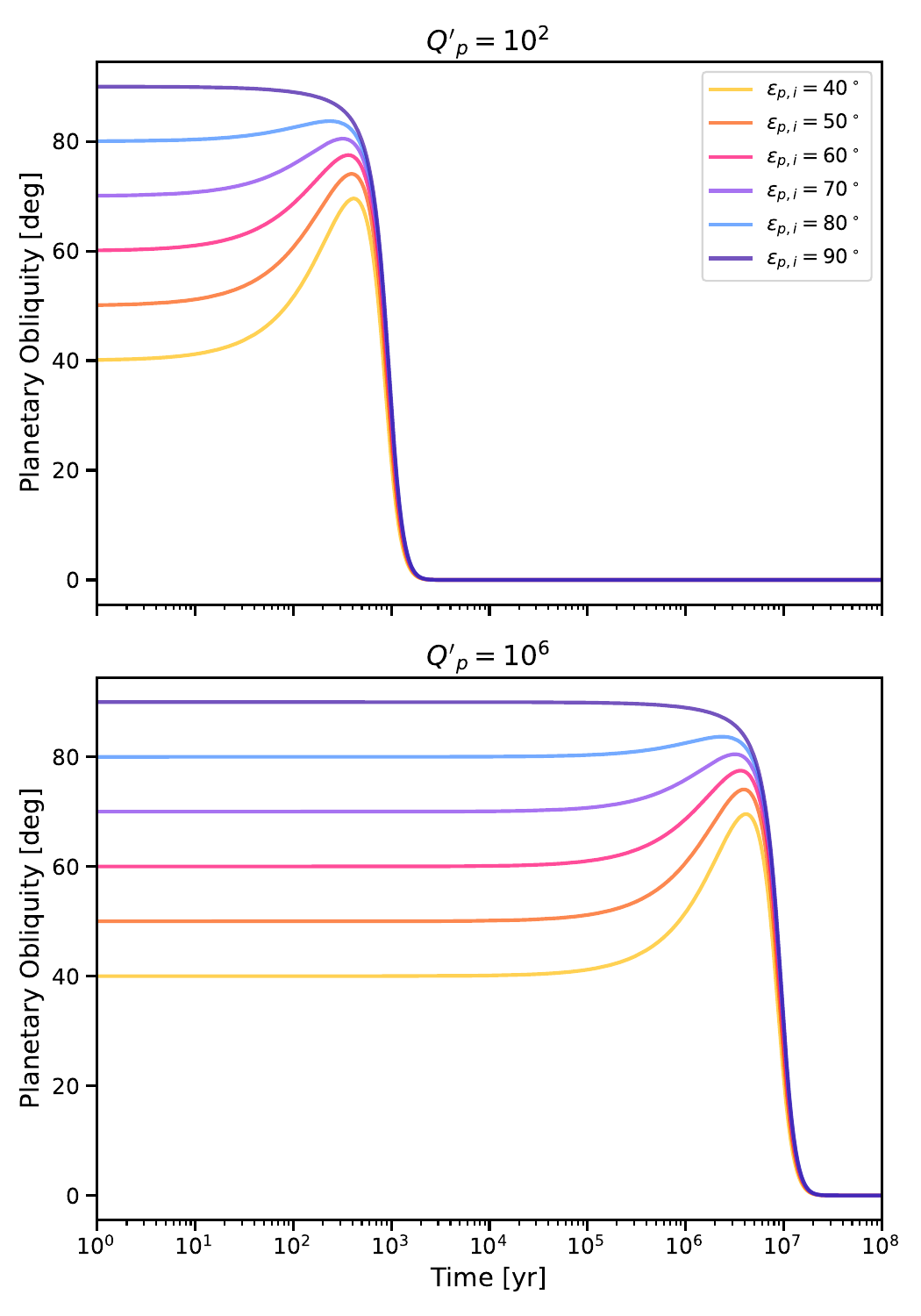}
    \caption{Time evolution of obliquity for $Q'_p = 10^2$ (top) and $Q'_p = 10^6$ (bottom). Curves are initialized at different obliquities, with colors indicating the initial values.}
    \label{fig:obliquity_evol}
\end{figure}

To evolve the system secularly, we solve the coupled evolution equations for $a$, $e$, planetary spin rate $\omega_p$, and $\epsilon_p$ given by \citep{2010A&A...516A..64L}:

\begin{align} \label{da_dt}
       \frac{1}{a}\frac{da}{dt} = \frac{4a}{GM_\star M_p}\Bigg\{ K_p\left [N(e)x_p\frac{\omega_p}{n} - N_a(e) \right]\Bigg\},
\end{align}
\begin{align} \label{de_dt}
       \frac{1}{e}\frac{de}{dt} = \frac{11a}{GM_\star M_p}\Bigg\{ K_p\left [\Omega_e(e)x_p\frac{\omega_p}{n} - \frac{18}{11}N_e(e) \right] \Bigg\},
\end{align}
\begin{align} \label{domega_dt}
       \frac{dC_p\omega_p}{dt} = -\frac{K_p}{n} \left[ (1 + x_p^2) \Omega(e) \frac{\omega_p}{n} - 2x_pN(e) \right] ,
\end{align}
\begin{align} \label{dep_dt}
       \frac{d\epsilon_p}{dt} = \sin{\epsilon_p}\frac{K_p}{C_p\omega_p n} \left[ (x_p - \eta_p) \Omega(e)\frac{\omega_p}{n} - 2N(e) \right] ,
\end{align}
where $x_p = \cos{\epsilon_p}$, and $C_p$ is the principal moment of inertia of the deformable planet. For V1298 Tau b, we adopt $C_p = 0.25M_pR_p^2$, as sub-Neptune to giant planets are expected to have moment of inertia coefficients of this order \citep{1999ssd..book.....M}. $\eta_p$ is the ratio between planet's rotational and angular momentum given by \citep{2010A&A...516A..64L} as:
\begin{equation}
    \eta_p = \frac{M_\star + M_p}{M_p M_\star}\frac{C_p \omega_p}{a^2n\sqrt{1 - e^2}}.
\end{equation}
$N_e (e), \Omega_e(e)$ are defined as:
\begin{equation} 
    N_e(e) = \frac{1 + \frac{15}{4}e^2 + \frac{15}{8}e^4 + \frac{5}{64}e^6}{(1-e^2)^{13/2}},
\end{equation}
\begin{equation}
    \Omega_e(e) = \frac{1 + \frac{3}{2}e^2 + \frac{1}{8}e^4}{(1 - e^2)^5}.
\end{equation}
Here we ignore the influence of tides raised by the planet on the star. We have verified that including stellar tides does not affect the results, as tides raised by the star on the planet dominate the evolution.

The coupled ODEs are initialized using the present-day orbital parameters of V1298 Tau b, its observed semi-major axis and eccentricity. Since the planetary rotation rate is unknown and young planets may rotate rapidly, we adopt an initial spin rate $\omega_p=2\pi/P_{\rm rot}$ with $P_{\rm rot}=10$ hr. This rapid initial rotation provides an upper limit on $\tau_{\rm damp}$, since slower initial rotation leads to faster obliquity damping. We evolve the system over a range of initial obliquities, $0^\circ<\epsilon_{p,i}<90^\circ$, and reduced tidal quality factors, $10\le Q'_p\le10^6$. Representative examples of the resulting obliquity evolution are shown in Figure~\ref{fig:obliquity_evol}.

\section{Constructing the Laplace Lagrange Nodal Matrix} \label{app:LL}
The off-diagonal elements of the Laplace–Lagrange nodal matrix are given by \citep{1999ssd..book.....M}:
\begin{equation}
    B_{jk} = +n_j \frac{1}{4}\frac{m_k}{M_\star + m_j} \alpha_{jk}\bar{\alpha}_{jk}b_{3/2}^{(1)}(\alpha_{jk}),
\end{equation}
where $j, k \in \{ 1, 2, 3, 4 \}$ index the planets in the V1298 Tau system. Here, $n_j$ is the mean motion of planet $j$, and $m_j$ and $m_k$ are the masses of planets $j$ and $k$, respectively. 
$\alpha_{jk} = a_j/a_k$ if $a_j < a_k$ and $\alpha_{jk} = a_k/a_j$ otherwise. $\bar{\alpha}_{jk} = \alpha_{jk}$ for $a_j < a_k$, and  $\bar{\alpha}_{jk} = 1$ otherwise. The function $b_{3/2}^{(1)}(\alpha_{jk})$
is the Laplace coefficient, defined as by
\begin{equation}
    b_{3/2}^{(1)}(\alpha_{jk}) = \frac{1}{\pi}\int_0^{2\pi} \frac{\cos{\psi}}{(1 - 2\alpha_{jk}\cos{\psi} + \alpha_{jk}^2)^{3/2}}d\psi.
\end{equation}
The diagonal elements of the matrix are given by the negative sum of the off-diagonal elements in each row,
\begin{equation}
    B_{jj} = -\sum_{k \ne j} B_{jk},
\end{equation}
which ensures conservation of angular momentum and reflects that the net torque on each planet is the sum of the torques exerted by all other planets in the system.

\bibliography{references.bib}
\bibliographystyle{aasjournal}

\restartappendixnumbering
\end{document}